\documentclass[english,conference,10pt]{IEEEtran}
\usepackage[T1]{fontenc}
\usepackage[latin9]{inputenc}
\usepackage{float}
\usepackage{amsthm}
\usepackage{amsmath}
\usepackage{amssymb}
\usepackage{graphicx}
\usepackage{esint}

\makeatletter

\floatstyle{ruled}
\newfloat{algorithm}{tbp}{loa}
\providecommand{\algorithmname}{Algorithm}
\floatname{algorithm}{\protect\algorithmname}

\theoremstyle{plain}
\newtheorem{thm}{\protect\theoremname}
\theoremstyle{plain}
\newtheorem{lem}[thm]{\protect\lemmaname}

\usepackage{amsmath}
\usepackage{mathabx}
\usepackage{stmaryrd}

\DeclareMathOperator{\enum}{e}
\DeclareMathOperator{\inum}{i}

\DeclareMathOperator{\sinc}{sinc}

\DeclareMathOperator{\Si}{Si}

\allowdisplaybreaks
\IEEEoverridecommandlockouts

\author{\IEEEauthorblockN{Sander Wahls}
\IEEEauthorblockA{Delft Center for Systems and Control \\ TU Delft, Delft, The Netherlands \\ Email: s.wahls@tudelft.nl}
\and
\IEEEauthorblockN{H. Vincent Poor}
\IEEEauthorblockA{Department of Electrical Engineering\\
Princeton University, Princeton, NJ, USA \\
Email: poor@princeton.edu}}

\makeatother

\usepackage{babel}
\providecommand{\lemmaname}{Lemma}
\providecommand{\theoremname}{Theorem}

\begin{document}

\title{Fast Inverse Nonlinear Fourier Transform For Generating Multi-Solitons
In Optical Fiber \thanks{This research was supported in part by the U.S. National Science Foundation under Grant CCF-1420575.}}
\maketitle
\begin{abstract}
The achievable data rates of current fiber-optic wavelength-division-multiplexing
(WDM) systems are limited by nonlinear interactions between different
subchannels. Recently, it was thus proposed to replace the conventional
Fourier transform in WDM systems with an appropriately defined nonlinear
Fourier transform (NFT). The computational complexity of NFTs is a
topic of current research. In this paper, a fast inverse NFT algorithm
for the important special case of multi-solitonic signals is presented.
The algorithm requires only $\mathcal{O}(D\log^{2}D)$ floating point
operations to compute $D$ samples of a multi-soliton. To the best
of our knowledge, this is the first algorithm for this problem with
$\log^{2}$-linear complexity. The paper also includes a many samples
analysis of the generated nonlinear Fourier spectra.
\end{abstract}

\section{Introduction}

The current generation of fiber-optic communication systems based
on wavelength-division-multiplexing (WDM) is operating close to capacity
\cite{Essiambre2010}. In order to avoid a looming capacity crunch,
new approaches beyond WDM need to be explored. The nonlinear effects
in optical fiber lead to interference between different subchannels,
and have been identified as one of the major capacity-limiting factors.
In order to avoid interference between subchannels, it was recently
proposed to modulate information using a nonlinear Fourier transform
(NFT) instead of the traditional Fourier transform used in WDM \cite{Yousefi2012a,Yousefi2012b,Yousefi2013}.
Analogously to how the spatial evolution of a signal in a linear channel
such as copper cable becomes trivial in the traditional Fourier domain,
the evolution of a signal in optical fiber becomes trivial in a properly
defined nonlinear Fourier domain. The most important aspect here is
that different regions in the nonlinear Fourier domain do not interact
as the signal travels through the fiber. Thus, by defining subchannels
in the nonlinear Fourier domain, interference among them can be avoided.
There are many interesting open questions regarding information transmission
in the nonlinear Fourier domain, with one of them being computational
complexity. While the celebrated fast Fourier transform has been the
workhorse of practical implementations of WDM, the first fast NFTs
have been proposed only recently \cite{Wahls2013b,Wahls2013d}. The
inverse NFT however, in contrast to the traditional case, differs
from the forward transform. No fast inverse NFT has been established
in the literature so far to the best of our knowledge, although inverse
NFTs are used in several practical problems \cite{Rourke1992,Brenne2003}.
The goal of this paper is to, at least partially, close this gap for
the important case of multi-solitonic signals. The algorithm developed
in this paper can generate multi-solitons with arbitrary discrete
spectra (but not norming constants). Such signals have recently been
considered for information transmission in \cite{Yousefi2013} and
\cite{Hari2014}.

The paper is structured as follows. In Section \ref{sec:Nonlinear-Fourier-Transform},
the continuous-time NFT is introduced and discretized. The proposed
algorithm will operate in two stages, which are then developed in
the Sections \ref{sec:Synthesis-of-Multi-Solitonic} and \ref{sec:Fast-Algorithm}.
Two numerical examples follow in Section \ref{sec:Numerical-Example}.
Section \ref{sec:Conclusions} finally concludes the paper.

\section{Nonlinear Fourier Transform\label{sec:Nonlinear-Fourier-Transform}}

In this section, the nonlinear Fourier transform is first established
for continuous-time signals. The NFT is then discretized and, finally,
the inverse discrete NFT is discussed.

\subsection{Continous-Time Transform}

The nonlinear Fourier transform {[}w.r.t. the signal model in Eq.
(\ref{eq:NSE}) below{]} is due to Zakharov and Shabat \cite{Zakharov1972}.
Let $\{q(t)\}_{t\in\mathbb{R}}\subset\mathbb{C}$ denote a vanishing
signal. Its NFT is found through the analysis of the associated scattering
problem 
\begin{align}
\frac{d}{dt}\boldsymbol{\mathbf{\phi}}(t,\lambda)= & \left[\begin{array}{cc}
-\inum\lambda & q(t)\\
-\bar{q}(t) & \inum\lambda
\end{array}\right]\boldsymbol{\mathbf{\phi}}(t,\lambda),\label{eq:Z-S-1}\\
\boldsymbol{\phi}(t,\lambda)= & \left[\begin{array}{c}
\enum^{-\inum\lambda t}\\
0
\end{array}\right]+o(1),\quad t\to-\infty.\label{eq:Z-S-2}
\end{align}
With $\phi_{1}$ and $\phi_{2}$ denoting the components of $\boldsymbol{\phi}$,
define
\begin{equation}
\alpha(\lambda):=\lim_{t\to\infty}\enum^{\inum\lambda t}\phi_{1}(t,\lambda),\:\beta(\lambda):=\lim_{t\to\infty}\enum^{-\inum\lambda t}\phi_{2}(t,\lambda),\label{eq:alpha-beta}
\end{equation}
where $\lambda\in\mathbb{C}$. The functions $\alpha$ and $\beta$
are usually not well-defined on the whole complex plane. With some
assumptions on the signal $q(t)$, however, the following can be said.
\begin{lem}
[\cite{Ablowitz1974}]If there exist positive constants $C$ and
$\kappa$ such that $|q(t)|\le C\enum^{-2\kappa|t|}$ for all real
$t$, then:
\begin{enumerate}
\item $\alpha(\lambda)$ is analytic in the closed upper half-plane $\Im(\lambda)\ge0$
\item $\alpha(\lambda)$ has finitely many roots $\lambda_{1},\dots,\lambda_{J}$
in $\Im(\lambda)\ge0$
\item $\beta(\lambda)$ is analytic around the real line $\mathbb{R}$ and
well-defined at the roots $\lambda_{1},\dots,\lambda_{J}$ of $\alpha(\lambda)$
in $\Im(\lambda)>0$
\end{enumerate}
\end{lem}

This lemma implies that the \emph{reflection coefficient} 
\begin{equation}
\hat{q}(\omega):=\frac{\beta(\omega)}{\alpha(\omega)},\quad\omega\in\mathbb{R},\label{eq:reflection-coefficient}
\end{equation}
is well-defined except at finitely many poles. The function $\beta/\alpha$
is also well-defined at the (complex) roots $\lambda_{j}$. The residual
of $\beta/\alpha$ at these points is, assuming that all $\lambda_{j}$
are simple roots, given by 
\[
\tilde{q}_{j}:=\beta(\lambda_{j})\Big/\frac{d\alpha}{d\lambda}(\lambda_{j}).
\]

With these definitions in mind, the \emph{nonlinear Fourier transform}
(or \emph{scattering data}) of $q(t)$ is defined as \cite[Apdx. 5]{Ablowitz1974}
\begin{equation}
\left\{ \hat{q}(\omega)\right\} _{\omega\in\mathbb{R}},\quad\left(\lambda_{j},\tilde{q}_{j}\right)_{j=1}^{J}.\label{eq:nonlinear-Fourier-spectrum}
\end{equation}
Please note that the nonlinear Fourier spectrum has a physical interpretation
\cite{Wahls2014a}. The reflection coefficient reduces to the linear
Fourier transform for signals with small amplitudes. It describes
the so-called \emph{radiation} components of the signal. The roots
$\lambda_{k}$ indicate \emph{solitonic components} in the signal,
which are further specified through the norming constants $\tilde{q}_{k}$.
(Solitons are basically ``travelling humps'' that have no analogue
in linear evolution equations.) A \emph{multi-soliton} is a signal
that contains several solitonic components. A signal is \emph{reflectionless}
if the reflection coefficient is zero everywhere.

The nonlinear Fourier transform has several properties similar to
the conventional Fourier transform \cite[Sec. IV.D]{Yousefi2012a}.
In particular, the effects of shifting or dilating the signal on the
nonlinear Fourier transform turn out to be simple:
\begin{align}
q(t-t_{0}) & \leftrightarrow\{\enum^{-2\inum\omega t_{0}}\hat{q}(\omega)\},\,(\lambda_{j},\enum^{-2\inum\lambda_{j}t_{0}}\tilde{q}_{j}),\label{eq:time-shift}\\
q(t/\eta) & \leftrightarrow\{\widehat{|\eta|q}(\eta\omega)\},\,(\eta\lambda_{j},\widetilde{|\eta|q}_{j}).\label{eq:time-dilation}
\end{align}
The most important property is however the following. Consider a space-time
signal $\{\mathcal{Q}(x,t)\}_{x\ge0,t\in\mathbb{R}}\subset\mathbb{C}$
that is governed by the \emph{focusing nonlinear Schr\"odinger equation
(NSE)} 
\begin{align}
\inum\frac{\partial\mathcal{Q}}{\partial x} & =\frac{\partial^{2}\mathcal{Q}}{\partial t^{2}}+|\mathcal{Q}|^{2}\mathcal{Q}.\label{eq:NSE}
\end{align}
Denote the functions $\alpha(z)$ and $\beta(z)$ that correspond
to the signal $q(t)=\mathcal{Q}(x,t)$ by $\alpha(x,\lambda)$ and
$\beta(x,\lambda)$. Then, \cite[V]{Yousefi2012a} 
\[
\alpha(x,\lambda)=\alpha(0,\lambda),\quad\beta(x,\lambda)=\enum^{-4\inum\lambda^{2}x}\beta(0,\lambda).
\]
The focusing NSE describes the evolution of a signal in lossless,
noise-free optical fiber \cite{Yousefi2012a}. The evolution of a
signal thus indeed becomes trivial in the nonlinear Fourier domain.

\subsection{Discrete-Time Transform}

In order to compute a numerical approximation of the nonlinear Fourier
spectrum (\ref{eq:nonlinear-Fourier-spectrum}), the signal $q(t)$
is considered only on a (sufficiently large) finite time interval.
Since the effects of time-shifting and dilation are simple {[}see
(\ref{eq:time-shift})--(\ref{eq:time-dilation}){]}, we can choose
the interval $[-1,0]$ in order to simplify the following exposition
without loss of generality. In this interval, $D$ equidistant scaled
samples
\[
Q[n]:=\epsilon q\left(-1+n\epsilon-\frac{\epsilon}{2}\right),\quad\epsilon:=\frac{1}{D},\quad n\in\{1,\dots,D\},
\]
are taken in order to represent to signal. It will be convenient to
transform the spectral parameter $\lambda$ as follows: 
\begin{equation}
z=z(\lambda):=\enum^{-2\inum\lambda\epsilon},\qquad\Re(\lambda)\in\left[-\frac{\pi}{2\epsilon},\frac{\pi}{2\epsilon}\right].\label{eq:coordinate-transform}
\end{equation}
(The real part of $\lambda$ has been constrained so that the inverse
transform $z\mapsto\lambda(z)$ is well-defined.) Under the assumption
that $q(t)$ is piecewise constant and that $\epsilon$ is infinitesimal,
the scattering problem (\ref{eq:Z-S-1})--(\ref{eq:Z-S-2}) reduces
to \cite[Sec. II]{Brenne2003} 
\begin{align}
\boldsymbol{\phi}[n,z]:= & \frac{z^{\frac{1}{2}}}{\sqrt{1+|Q[n]|^{2}}}\left[\begin{array}{cc}
1 & z^{-1}Q[n]\\
-\bar{Q}[n] & z^{-1}
\end{array}\right]\boldsymbol{\phi}[n-1,z],\label{eq:discrete-Z-S-1}\\
\boldsymbol{\phi}[0,z]:= & z^{-\frac{D}{2}}\left[\begin{array}{c}
1\\
0
\end{array}\right].\label{eq:discrete-Z-S-2}
\end{align}
Here, the initial condition (\ref{eq:discrete-Z-S-2}) was obtained
by evaluating (\ref{eq:Z-S-2}) at $t=-1$. Similarly, after considering
(\ref{eq:alpha-beta}) at $t=0$, the functions $\alpha(\lambda)$
and $\beta(\lambda)$ are approximated by
\begin{equation}
a(z):=\phi_{1}[D,z],\quad b(z):=\phi_{1}[D,z].\label{eq:a(z)-b(z)}
\end{equation}

Note that the complex half-strip $\Re(\lambda)\in[-\frac{\pi}{2\epsilon},\frac{\pi}{2\epsilon}]$,
$\Im(\lambda)>0$ with respect to the original spectral parameter
$\lambda$ corresponds to the exterior of the unit circle $|z|>1$.
The real interval $[-\frac{\pi}{2\epsilon},\frac{\pi}{2\epsilon}]$,
in particular, is mapped to the unit circle $|z|=1$. The reflection
coefficient $\hat{q}(\omega)$ is therefore approximated as
\begin{equation}
\hat{Q}(\omega):=\frac{b(z(\omega))}{a(z(\omega))},\quad\omega\in\left[-\frac{\pi}{2\epsilon},\frac{\pi}{2\epsilon}\right].\label{eq:discrete-reflection-coefficient}
\end{equation}
The roots $z_{1},\dots,z_{K}$ of $a(z)$ outside the unit circle
$|z|>1$ are taken as approximations of the roots $\lambda_{1},\dots,\lambda_{J}$
of $\alpha(\lambda)$ subject to the coordinate transform (\ref{eq:coordinate-transform}).
Note that there may be spurious roots $z_{k}$ that do no correspond
to any of the $\lambda_{j}$. These spurious roots are numerical artifacts.
They usually occur in physically unplausible regions (e.g., very close
to or very far away from the unit circle) so that they can be easily
filtered out. The norming constants $\tilde{q}_{j}$ are approximated
through
\begin{equation}
\tilde{Q}_{k}:=b(z_{k})\Big/\frac{da}{dz}\frac{dz}{d\lambda}(\lambda_{k})=-\frac{b(z_{k})}{2\inum\epsilon z_{k}}\Big/\frac{da}{dz}(z_{k}).\label{eq:discrete-norming-constants}
\end{equation}
The \emph{discrete nonlinear Fourier transform} is then defined by
\begin{equation}
\{\hat{Q}(\omega)\}_{\omega\in[-\frac{\pi}{2\epsilon},\frac{\pi}{2\epsilon}]},\quad\{z_{k},\tilde{Q}_{k}\}_{k=1}^{K}.\label{eq:discrete-nonlinear-Fourier-spectrum}
\end{equation}

\subsection{Inverse Discrete-Time Transform}

A close look at the finite difference equation (\ref{eq:discrete-Z-S-1})
reveals that $a(z)$ and $b(z)$ defined in (\ref{eq:a(z)-b(z)})
are neccessarily polynomials of $z^{-1}$ with a degree of at most
$D$ \cite[(19)]{Brenne2003}: 
\begin{equation}
a(z)={\textstyle \sum_{i=0}^{D-1}}a_{i}z^{-i},\; b(z)={\textstyle \sum_{i=0}^{D-1}}b_{i}z^{-i}.\label{eq:a-and-b-as-poly}
\end{equation}
Please note that the additional factors $z^{D/2}$ given in \cite[(19)]{Brenne2003}
cancel with the factor in (\ref{eq:discrete-Z-S-2}) in our case.
The following lemma specifies the range of the discrete nonlinear
Fourier transform.
\begin{lem}
[\cite{Brenne2003}] \label{lem:Brenne}Fix two arbirary polynomials
as in (\ref{eq:a-and-b-as-poly}). Then, there exists samples $Q[1],\dots,Q[D]$
such that (\ref{eq:a(z)-b(z)}) is satisfied if and only $a_{0}\ge0$
and $|a(\xi)|^{2}+|b(\xi)|^{2}=1$ for all $|\xi|=1$. In that case,
$a_{0}=1/\prod_{n=1}^{D}\sqrt{1+|Q[n]|^{2}}>0$ and 
\begin{equation}
Q[D]=\frac{a_{D-1}}{b_{D-1}}=-\frac{\bar{b}_{0}}{\bar{a}_{0}}.\label{eq:Q[D]}
\end{equation}
\end{lem}
\begin{IEEEproof}
The first part in shown in \cite[Sec. IV+Apdx.]{Brenne2003}. The
other two formulas are given in \cite[Eqs. (14a)+(15)+(37)]{Brenne2003}.
The last formula deviates from \cite{Brenne2003} because here $Q[n]=-\bar{\rho}_{n}$.
\end{IEEEproof}
There are various ways to approach the inverse nonlinear Fourier transform.
See, e.g., \cite{Yousefi2013}. The fast  $\mathcal{O}(D\log^{2}D)$
algorithm that will be proposed in the following two sections uses
a two-step procedure similar to \cite{Brenne2003} and \cite{Wahls2014a}:
\begin{enumerate}
\item Synthesize $a(z)$ and $b(z)$ such that the conditions in Lemma \ref{lem:Brenne}
are met and the roots of $a(z)$ are $z_{k}:=z(\lambda_{k})$.
\item Recover the samples $Q[n]$ from (\ref{eq:discrete-Z-S-1})--(\ref{eq:discrete-Z-S-2}).
\end{enumerate}
The complexities of comparable algorithms are higher: either $\mathcal{O}(D^{2})$,
as in \cite{Wahls2014a} and \cite{Brenne2003}; $\mathcal{O}(DK^{2})$
as in \cite{Rourke1992} and \cite[III.B-3]{Yousefi2013}; or even
$\mathcal{O}(D^{3})$ as in \cite[III.B-1]{Yousefi2013}. Here $K$
is the number of solitons. (Note that $K\sim D$ in communication
problems.)

\section{Step 1: Fast Generation of $a(z)$ and $b(z)$ \label{sec:Synthesis-of-Multi-Solitonic}}

In this section, the first step of our proposed algorithm is described.
The basic approach used here is due to our previous work \cite{Wahls2014a}.
However, the construction given here avoids a pathology from which
the construction in \cite{Wahls2014a} suffers, as a new many-samples
analysis presented below shows. Please note that no complexity analysis
of this step will be given here because the analysis given in \cite[VI]{Wahls2014a}
still applies.

\subsection{Basic Construction}

Consider any polynomial $\psi(z)=\sum_{i=0}^{D-1}\psi_{i}z^{-i}$
that fulfills 
\[
|\psi(z)|\in[0,1]\qquad\forall|z|=1.
\]
Through spectral factorization \cite{Sayed2001}, two polynomials
$u(z)=\sum_{i=0}^{D-1}u_{i}z^{-i}$ and $b(z)=\sum_{i=0}^{D-1}b_{i}z^{-i}$
can be found that have no roots outside the unit circle and satisfiy
\begin{equation}
|u(\xi)|^{2}=1-\delta^{2}|\psi(\xi)|^{2},\,|b(\xi)|^{2}=\delta^{2}|\psi(\xi)|^{2}\quad\forall|\xi|=1.\label{eq:prop-b}
\end{equation}
Here, $\delta\in(0,1)$ is a free parameter. We choose 
\begin{equation}
a_{\text{ideal}}(z):=u(z)\prod_{k=1}^{K}\frac{z-z_{k}}{1-z\bar{z}_{k}}.\label{eq:a-ideal}
\end{equation}
This ideal $a_{\text{ideal}}(z)$ has the right roots in $|z|>1$
and satisfies
\begin{align}
|a_{\text{ideal}}(\xi)|^{2}+|b(\xi)|^{2}= & |u(\xi)|^{2}\prod_{k=1}^{K}|\bar{\xi}|\frac{|\xi-z_{k}|}{|\overline{\xi-z_{k}}|}+|b(\xi)|^{2}\nonumber \\
(\text{use }|\xi|=1)\quad= & 1-\delta^{2}|\psi(\xi)|^{2}+\delta^{2}|\psi(\xi)|^{2}=1,\label{eq:|aideal|+|b|}
\end{align}
for all $|\xi|=1$, but it is not a polynomial in $z^{-1}$, in general.

However, since $a_{\text{ideal}}(z)$ is rational with all poles inside
the unit circle, it has an expansion $a_{\text{ideal}}(z)=\sum_{i=0}^{\infty}a_{\text{ideal},i}z^{-i}$
in $|z|\ge1$ with coefficients that vanish exponentially vast. Thus,
$a_{\text{ideal}}(z)$ can be easily approximated by a polynomial
$p(z)=\sum_{i=0}^{D-1}p_{i}z^{-i}$ up to any desired precision given
that $D$ is not too small. We shall assume this in the following,
and set
\[
a(z):=\enum^{\inum\varphi}{\textstyle \sum_{i=0}^{D-1}}p_{i}z^{-i}=\enum^{\inum\varphi}a_{\text{ideal}}(z)+\mathcal{O}(\epsilon^{2}),
\]
where $\varphi\in[-\pi,\pi]$ is such that $a(z)$ fulfills the condition
$\lim_{z\to\infty}a(z)\ge0$ in Lemma \ref{lem:Brenne}. In light
of (\ref{eq:|aideal|+|b|}), one finds that $a(z)$ and $b(z)$ also
fulfill the other condition in Lemma \ref{lem:Brenne} (up to a rapidly
vanishing error). Thus, the problem of recovering $Q[1],\dots,Q[D]$
through (\ref{eq:discrete-Z-S-1})--(\ref{eq:discrete-Z-S-2}) is
well-defined.

\subsection{Choice Of The Filter $\psi(z)$}

In our previous paper \cite{Wahls2014a}, the filter $\psi(z)=z^{-(D-1)}$
and a very small $\delta$ were used. However, an analysis similar
to the one below shows that this choice results in a reflection coefficient
that does not vanish in the many samples regime for $\omega\to\pm\infty$.
Parseval's identity for the NFT \cite[p. 4320]{Yousefi2012a} however
implies that any relection coefficent of a signal with finite energy
must vanish. Although this can be fixed by letting $\delta$ depend
on $\epsilon$, one then obtains the problem that the norming constants
converge towards zero as $D\to\infty$. Again, this cannot happen
for any real signal. In order to avoid these two problems, one may
fix $\delta$ independently of $\epsilon$ and use a proper low-pass
filter for $\psi(z)$ to ensure that the reflection coefficient vanishes
at higher frequencies. In our numerical experiments, a simple truncated
ideal low-pass was used \cite[p. 98]{Papoulis1962}: 
\begin{equation}
\psi(z):=\sum_{i=0}^{D-1}\frac{2\omega_{c}}{D\pi}\sinc\left(\frac{2\omega_{c}}{D\pi}\left(i-\frac{D-1}{2}\right)\right)z^{-i},\label{eq:truncated-low-pass}
\end{equation}
where $\omega_{c}=10$ is the desired cut-off frequency for the reflection
coefficient (\ref{eq:reflection-coefficient}). Please note that the
filter $\psi(z)$ is independent of $q(t)$ and thus needs to be designed
only once.

\subsection{Many Samples Analysis\label{sub:Many-Samples-Analysis}}

In this subsection, the continous-time nonlinear Fourier spectrum
(\ref{eq:nonlinear-Fourier-spectrum}) that is generated through $a(z)$
and $b(z)$ introduced above is analyzed in the many samples regime
$D=1/\epsilon\to\infty$. The function $a(z)$ is constructed such
that its roots in $|z|>1$ match exactly the prespecified $\lambda_{1},\dots,\lambda_{K}$
under the transform $\lambda\mapsto z(\lambda)$. The behavior of
the reflection coefficient can be analyzed using (\ref{eq:discrete-reflection-coefficient}),
(\ref{eq:prop-b}) and (\ref{eq:|aideal|+|b|}): 
\begin{equation}
|\hat{q}(\omega)|^{2}=\lim_{D\to\infty}|\hat{Q}(\omega)|^{2}=\frac{\delta^{2}|\Psi(\omega)|^{2}}{1-\delta^{2}|\Psi(\omega)|^{2}},\quad\omega\in\mathbb{R}.\label{eq:|q(omega)|^2}
\end{equation}
Here, $|\Psi|$ denotes the asymptotic amplitude of (\ref{eq:truncated-low-pass})
given by 
\[
|\Psi(\omega)|:=\lim_{D\to\infty}|\psi(z(\omega))|=\frac{|\Si(\omega+\omega_{c})-\Si(\omega-\omega_{c})|}{\pi},
\]
where $\Si(s):=\int_{0}^{s}\frac{\sin\theta}{\theta}d\theta$
is the \emph{sine integral function} \cite[p. 98]{Papoulis1962}.
It now follows from $|\Psi(\omega)|\to\pm\frac{\pi}{2}$ that $q(\omega)\to0$
for $\omega\to\pm\infty$. That is, the pathology of the construction
in \cite{Wahls2014a} discussed in the previous subsection has been
fixed with the new design. The norming constants (\ref{eq:discrete-norming-constants})
are more difficult to analyze. In the appendix, they are shown to
satisfy 
\begin{equation}
\tilde{q}_{k}=\lim_{\epsilon\to0}\tilde{Q}_{k}=-\frac{B(\lambda_{k})}{U(\lambda_{k})}(\lambda_{k}-\bar{\lambda}_{k})\prod_{i\ne k}-\frac{\lambda_{k}-\lambda_{i}}{\lambda_{k}-\bar{\lambda}_{i}},\label{eq:asymptotic-norming-constants}
\end{equation}
where $B(\lambda):=\lim_{\epsilon\to0}b(z(\lambda))$ and $U(\lambda):=\lim_{\epsilon\to0}u(z(\lambda))$.
This formula shows that the norming constants have finite non-zero
limits in the many samples regime unless one of the prespecified $\lambda_{k}$
coincides with a root of either $B(\lambda_{k})$ and $U(\lambda_{k})$.
We have not yet performend an exact analysis of $B(\lambda)$ and
$U(\lambda)$. But since the functions $B(\lambda)$ and $U(\lambda)$
(and therefore their roots) are independent of the $\lambda_{1},\dots,\lambda_{K}$,
it seems unlikely that the pathology of vanishing norming constants
discussed above occurs with the new construction.

\begin{figure*}
\begin{centering}
\hfill{}\includegraphics[width=0.4\textwidth]{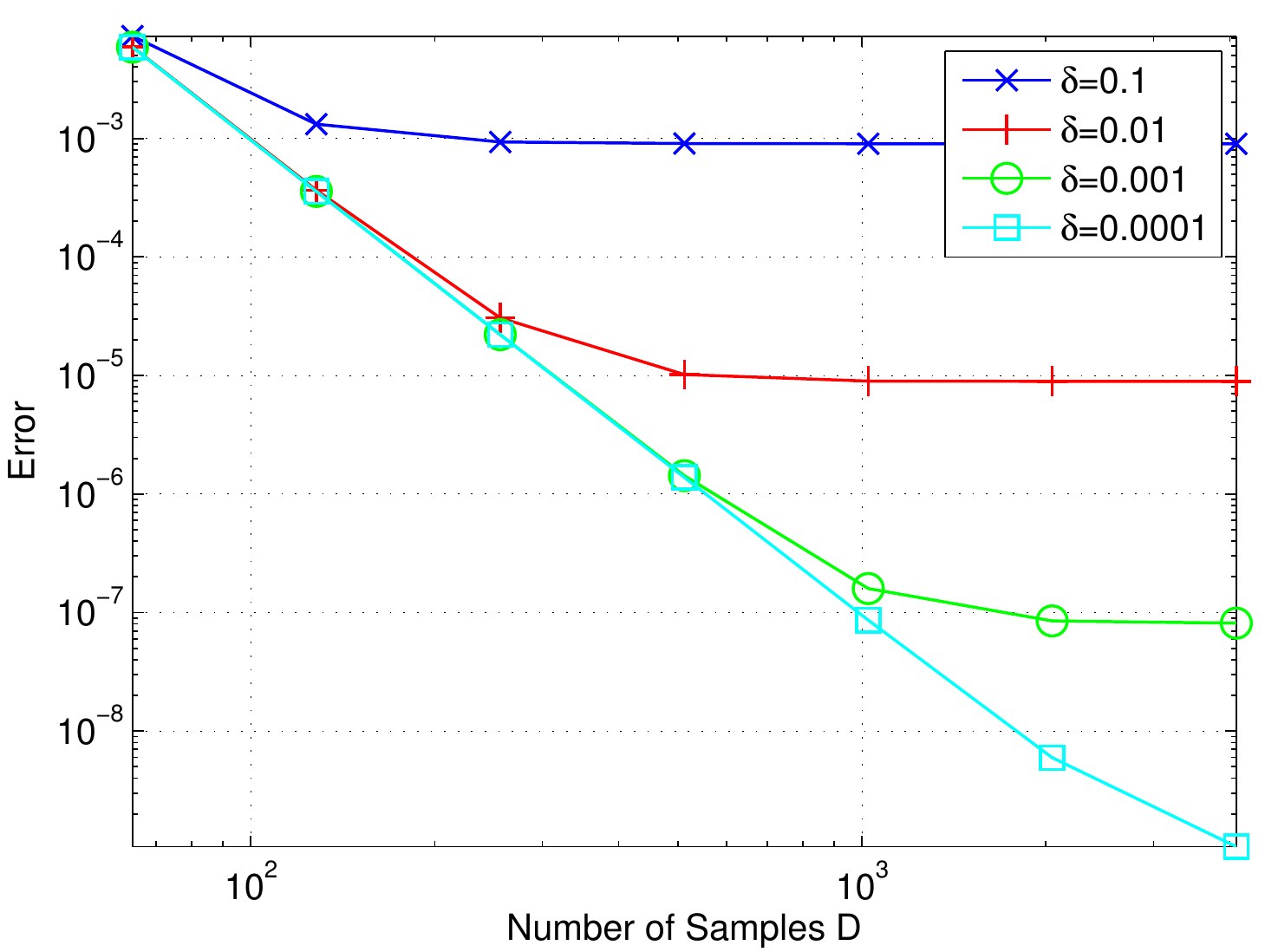}\hfill{}\includegraphics[width=0.4\textwidth]{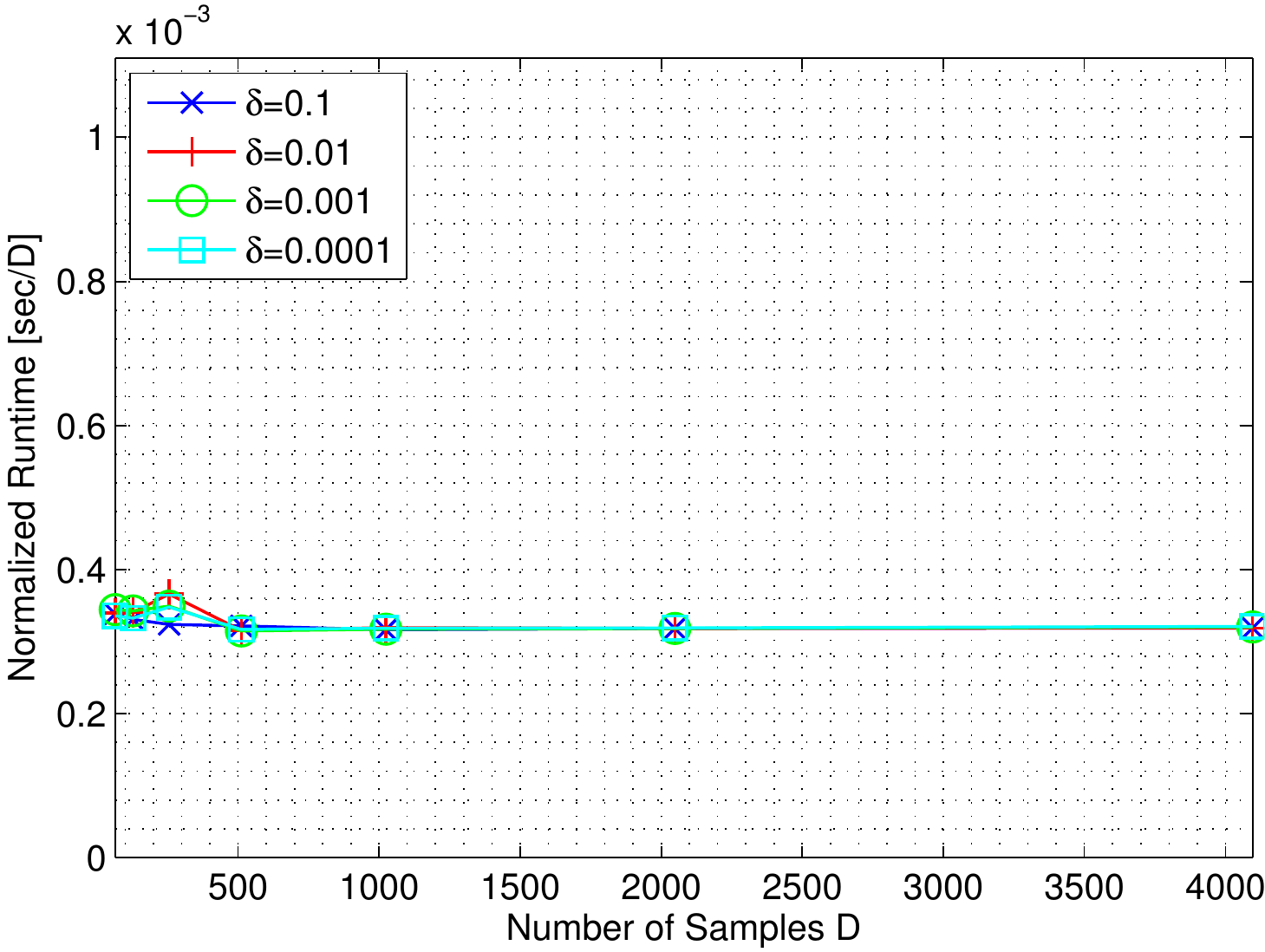}\hfill{}
\par\end{centering}

\caption{\label{fig:1st-example}First example. Left: Squared error $\sum_{n=1}^{D}|Q[n]-Q_{\text{exact}}[n]|^{2}/\sum|Q_{\text{exact}}[n]|^{2}$.
Right: Runtime per sample.}
\end{figure*}

\begin{figure*}
\begin{centering}
\hfill{}\includegraphics[width=0.4\textwidth]{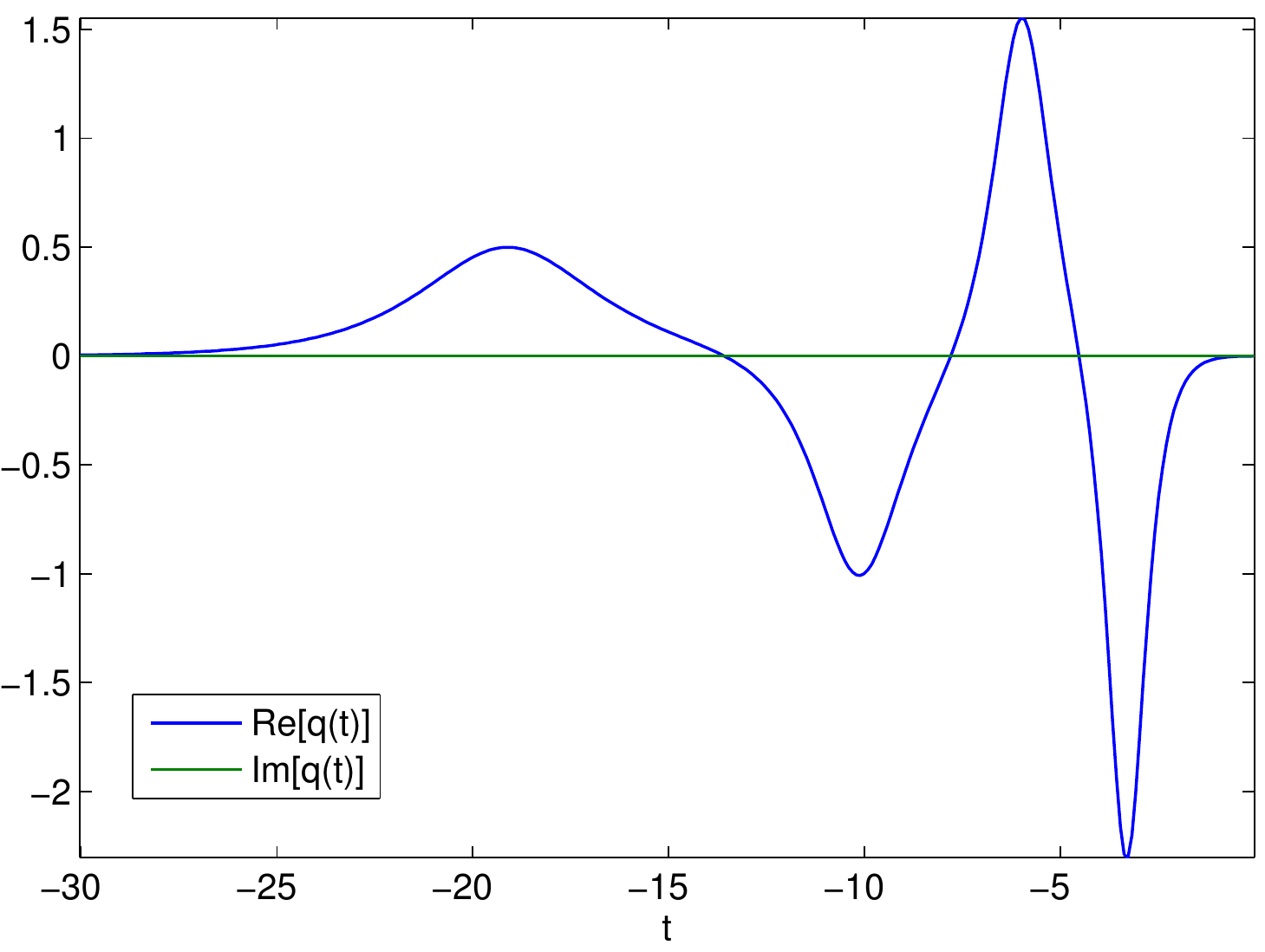}\hfill{}\includegraphics[width=0.4\textwidth]{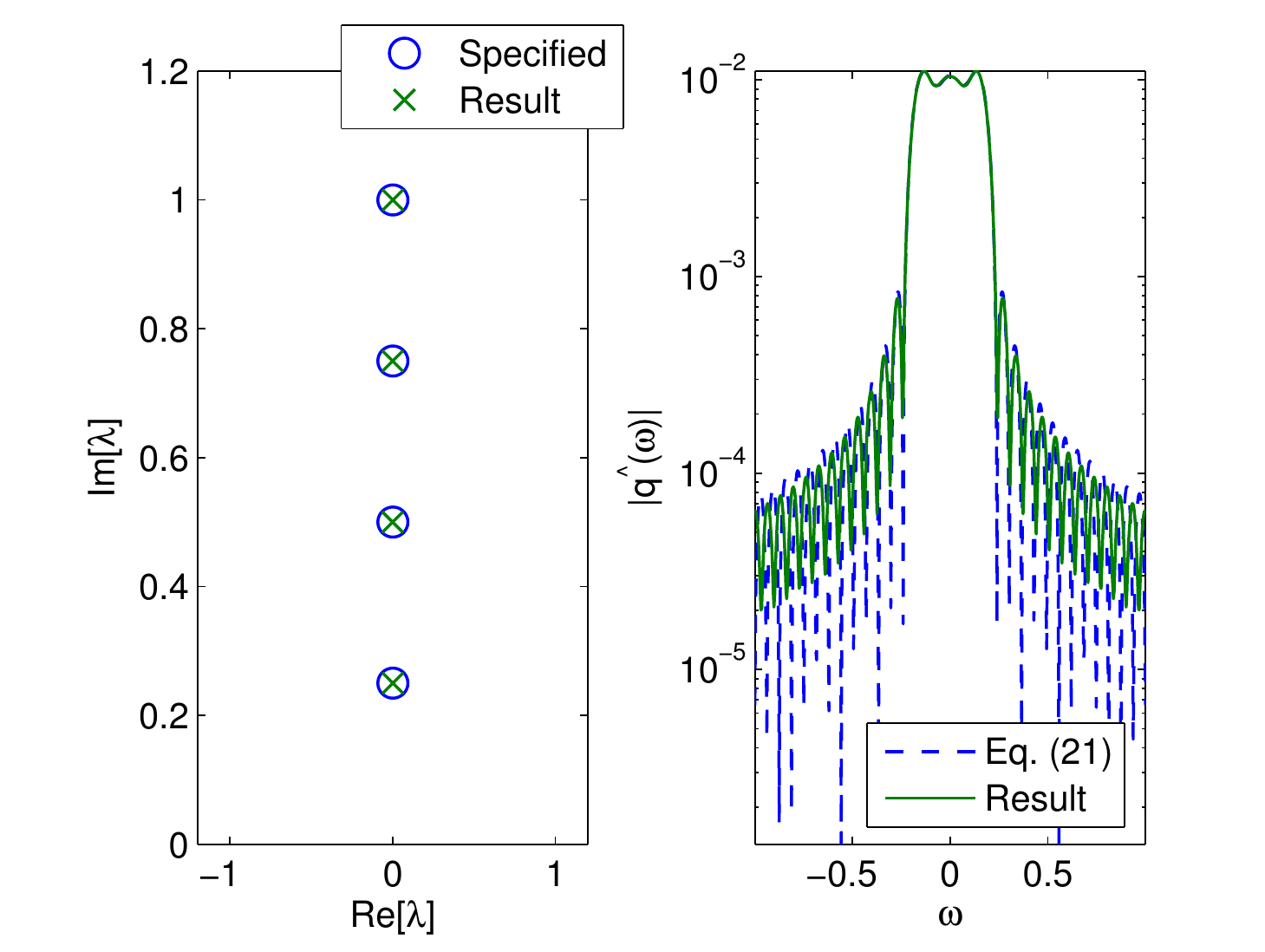}\hfill{}
\par\end{centering}

\caption{\label{fig:Second-example}Second example. Left: Generated time-domain
signal ($D=512$, $\delta=0.01$). Right: Roots of $a(z(\lambda))$
and reflection coefficient.}
\end{figure*}

\section{Step 2: Fast Inverse Scattering \label{sec:Fast-Algorithm}}

In this section, $a(z)$ and $b(z)$ are now considered known and
the samples $Q[1],\dots,Q[D]$ have to be recovered. A fast inverse
scattering algorithm that recovers $Q[1],\dots,Q[D]$ using only $\mathcal{O}(D\log^{2}D)$
floating point operations (\emph{flops}) for systems of the form (\ref{eq:discrete-Z-S-1})
has been proposed by McClary \cite[(2)]{McClary1983}. Unfortunately,
his algorithm does not work in our setup. In \cite{McClary1983},
the functions $a(z)$ and $b(z)$ are interpreted as $z$-transforms
of a down-going wave, which is sent into the earth, and a resulting
upcoming wave, which is measured at the same location. The initial
condition (\ref{eq:discrete-Z-S-2}) corresponds to the situation
where a unit pulse is sent into the earth and no upcoming wave is
generated, leading to a trivial medium that does not reflect (i.e.,
$Q[1]=\dots=Q[D]=0$). 

In the following, we will describe how McClary's approach can be modified
such that it works in our setup. By inverting (\ref{eq:discrete-Z-S-1})
and taking the definition (\ref{eq:a(z)-b(z)}) of $a(z)$ and $b(z)$
into account, one arrives at the following scattering problem: 
\begin{eqnarray}
\left[\begin{array}{c}
a_{n-1}(z)\\
b_{n-1}(z)
\end{array}\right] & = & \frac{z^{-\frac{1}{2}}}{\theta[n]}\left[\begin{array}{cc}
1 & -Q[n]\\
z\bar{Q}[n] & z
\end{array}\right]\left[\begin{array}{c}
a_{n}(z)\\
b_{n}(z)
\end{array}\right]\label{eq:inverse-discrete-Z-S-1}\\
\left[\begin{array}{c}
a_{D}(z)\\
b_{D}(z)
\end{array}\right] & = & \left[\begin{array}{c}
a(z)\\
b(z)
\end{array}\right],\qquad\theta[n]:=\sqrt{1+|Q[n]|^{2}}.\nonumber 
\end{eqnarray}
Assume that $D$ is even and split 
\[
a(z)=a^{L}(z)+z^{-\frac{D}{2}}a^{U}(z),\quad a^{L}(z):={\textstyle \sum_{i=0}^{\frac{D}{2}-1}}a_{i}z^{-i}.
\]
Split $b(z)=b^{L}(z)+z^{-\frac{D}{2}}b^{^{U}}(z)$ in the same way,
and define
\begin{align}
\mathbf{T}_{n}(z):= & \frac{z^{-\frac{1}{2}}}{\sqrt{1+|Q[n]|^{2}}}\left[\begin{array}{cc}
1 & -Q[n]\\
z\bar{Q}[n] & z
\end{array}\right],\label{eq:Tn}\\
\mathbf{T}_{m\to n}(z):= & \mathbf{T}_{n+1}(z)\mathbf{T}_{n+2}(z)\times\dots\times\mathbf{T}_{m}(z).\nonumber 
\end{align}
With this notation, one has that 
\begin{equation}
\left[\begin{array}{c}
a_{D-m}\\
b_{D-m}
\end{array}\right]=\mathbf{T}_{D\to m}\left[\begin{array}{c}
a^{L}\\
b^{L}
\end{array}\right]+z^{-\frac{D}{2}}\mathbf{T}_{D\to m}\left[\begin{array}{c}
a^{U}\\
b^{U}
\end{array}\right].\label{eq:Tmab}
\end{equation}
This equation leads to the important conclusion that the highest coefficient
of $a_{D-m}$ and $b_{D-m}$ is independent of $a_{U}$ and $b_{U}$
as long as $m<\frac{D}{2}$ (because $z^{-\frac{D}{2}}\mathbf{T}_{D\to m}$
will always shift the coefficients of $a_{U}$ and $b_{U}$ up in
that case.) Since $Q[D-m]$ can be recovered from the highest coefficients
through (\ref{eq:Q[D]}), this shows that the iteration (\ref{eq:inverse-discrete-Z-S-1})
with initial condition 
\[
a_{\frac{D}{2}}(z)=a^{U}(z),\quad b_{\frac{D}{2}}(z)=b^{U}(z)
\]
can be used to recover $Q[D],\dots,Q[\frac{D}{2}+1]$. This insight
ensures that the divide-and-conquer strategy given in Algorithm \ref{alg:Fast-inverse-scattering}
indeed recovers $Q[1],\dots,Q[D]$. The computational bottleneck in
Algorithm \ref{alg:Fast-inverse-scattering} are the polynomial products.
By applying the same complexity analysis that was used for \cite[Alg. 1]{Wahls2013d}
to the recursion tree generated by Algorithm \ref{alg:Fast-inverse-scattering},
one finds that its overall complexity is $\mathcal{O}(D\log^{2}D)$
flops if the fast Fourier transform is used to multiply polynomials
efficiently.

\begin{algorithm}
\textbf{\emph{Input:}} $a(z)=\sum_{i=0}^{D-1}a_{i}z^{-i}$ and $b(z)=\sum_{i=0}^{D-1}a_{i}z^{-i}$

\textbf{\emph{Output:}} $Q[1],\dots,Q[D]$ and $\mathbf{T}_{D\to1}(z)$

\textbf{if} $D=1$:
\begin{itemize}
\item \textbf{return} $Q[1]=a_{0}/b_{0}$ and $\mathbf{T}_{1}(z)$ in (\ref{eq:Tn})
\end{itemize}
\textbf{else}:
\begin{itemize}
\item call self with inputs $a^{U}(z)$ and $b^{U}(z)$ to recover $Q[D],\dots,Q[\frac{D}{2}+1]$
and $\mathbf{T}_{D\to D/2}(z)$
\item recover $a_{D/2}(z)$ and $b_{D/2}(z)$ via (\ref{eq:Tmab})
\item call self with inputs $a_{D/2}(z)$ and $b_{D/2}(z)$ to recover $Q[D/2],\dots,Q[1]$
and $\mathbf{T}_{D/2\to1}$
\item \textbf{return} $Q[1],\dots,Q[D]$ and $\mathbf{T}_{D\to D/2}\mathbf{T}_{D/2\to1}$
\end{itemize}
\caption{\label{alg:Fast-inverse-scattering}Recursive inverse scattering (requires
$D=2^{d}$)}
\end{algorithm}

\section{Numerical Examples \label{sec:Numerical-Example}}

In the first example, a single soliton with a root at $\lambda_{1}=20\inum$
is considered. For small $\delta>0$, the generated signal will be
almost reflectionless. For several number of samples $D$ and $\delta$'s,
the two steps of our algorithm as described in Sections \ref{sec:Synthesis-of-Multi-Solitonic}
and \ref{sec:Fast-Algorithm} have been carried out. Then, the forward
NFT was used to compute the norming constants of the generated signal.
Reflectionless signals with one soliton have a well-known closed-form
solution (e.g., \cite[II.D]{Wahls2014a}). We used this analytical
solution to generate samples $Q_{\text{exact}}[n]$ that represent
the exact one-soliton with the prespecified $\lambda_{1}$ and the
numerically determined norming constant. The error between the generated
signal and the analytical solution is depicted in Figure \ref{fig:1st-example}
(left). It eventually saturates because the generated signal is not
exactly reflectionless, but the error floors decrease as $\delta$
is decreased. The normalized runtimes are presented in Figure \ref{fig:1st-example}
(right). The runtime per sample grow only very slowly with $D$, confirming
the overall $\mathcal{O}(D\log^{2}D)$ runtime. (The runtimes do not
include the generation of $u(z)$ and $B(z)$ because they are independent
of the signal and have to be computed only once.)

The second example considers a problem from \cite{Hari2014}. A multi-soliton
with roots at $\lambda_{k}=0.25\inum k$, $k\in\{1,\dots,4\}$, is
generated. The roots were scaled by a factor $\eta=50$ such that
the signal would fit in the interval $[-1,0]$ (see (\ref{eq:time-dilation})).
The generated signal was rescaled to the original time-scale and analyzed
using the forward NFT. The results are depicted in Figure \ref{fig:Second-example}.
The forward NFT found the roots $\lambda_{k}$ of the generated signal
exactly in the right places. The reflection coefficient looks as predicted
in (\ref{eq:|q(omega)|^2}). It is interesting to compared Figure
\ref{fig:Second-example} (left) with \cite[Fig. 2b]{Hari2014}. It
was expected that the signals are different due to different norming
constants. However, our signal has approximately the same duration,
but with only half as many extrema. The highest peak is also about
only half as high as in \cite[Fig. 2b]{Hari2014}. We speculate that
the roots in \cite[Fig. 2b]{Hari2014} may be double because these
quantitative differences vanish if the orders of the roots are doubled
in our algorithm.

\section{Conclusions \label{sec:Conclusions}}

The (to the best of our knowledge) first fast inverse nonlinear Fourier
transform for the generation of multi-solitons has been presented.
The generated nonlinear Fourier spectra have been analyzed through
a many samples analysis. The $\mathcal{O}(D\log^{2}D)$ runtime and
accuracy of the algorithm have also been confirmend experimentally.
The current algorithm offers only rough control over the reflection
coefficient and norming constants, which should be addressed in future
research.

\bibliographystyle{IEEEtran}
\bibliography{isit15.bbl}

\newpage{}

\section*{Appendix: Derivation Of Equation (\ref{eq:asymptotic-norming-constants})}

The transformed coordinate $z=\enum^{-2\inum\lambda\epsilon}$ will
be close to one in the many samples regime $\epsilon=1/D\to0$. The
usual first order approximation of the exponential around zero,
\[
\enum^{s}=\sum_{i=0}^{\infty}\frac{s^{i}}{i!}=1+s+\mathcal{O}(|s|^{2}),
\]
shows that 
\[
z=1-2\inum\lambda\epsilon+\mathcal{O}(\epsilon^{2})
\]
and
\[
\bar{z}=\enum^{2\inum\bar{\lambda}\epsilon}=1+2\inum\bar{\lambda}\epsilon+\mathcal{O}(\epsilon^{2}).
\]
From these two formulas, one finds that 
\begin{align*}
\frac{z-z_{i}}{1-z\bar{z}_{i}}= & \frac{(1-2\inum\lambda\epsilon+\mathcal{O}(\epsilon^{2}))-(1-2\inum\lambda_{i}\epsilon+\mathcal{O}(\epsilon^{2}))}{1-(1-2\inum\lambda\epsilon+\mathcal{O}(\epsilon^{2}))(1+2\inum\bar{\lambda}_{i}\epsilon+\mathcal{O}(\epsilon^{2}))}\\
= & -\frac{\lambda-\lambda_{i}+\mathcal{O}(\epsilon^{2})}{\lambda-\bar{\lambda}_{i}+\mathcal{O}(\epsilon^{2})}
\end{align*}
and
\begin{align*}
\frac{1}{1-z_{k}\bar{z}_{k}}= & \frac{1}{1-(1-2\inum\lambda_{k}\epsilon+\mathcal{O}(\epsilon^{2}))(1+2\inum\bar{\lambda}_{k}\epsilon+\mathcal{O}(\epsilon^{2}))}\\
= & \frac{1}{2\inum\epsilon}\frac{1}{\lambda_{k}-\bar{\lambda}_{k}+\mathcal{O}(\epsilon^{2})}.
\end{align*}
The derivative of the Blaschke product 
\[
\mathcal{B}(z):=\prod_{i=1}^{K}\frac{z-z_{i}}{1-z\bar{z}_{i}}
\]
that occurs in the definition (\ref{eq:a-ideal}) of $a_{\text{ideal}}(z)$
is given by 
\[
\frac{d\mathcal{B}}{dz}(z)=\sum_{k=1}^{K}\frac{1-|z_{k}|^{2}}{(1-z\bar{z}_{k})^{2}}\prod_{i\ne k}\frac{z-z_{i}}{1-z\bar{z}_{i}},
\]
and satisfies
\[
\frac{d\mathcal{B}}{dz}(z_{k})=\frac{1}{1-|z_{k}|^{2}}\prod_{i\ne k}\frac{z_{k}-z_{i}}{1-z_{k}\bar{z}_{i}}.
\]
The approximations given above therefore imply that
\[
\frac{d\mathcal{B}}{dz}(z_{k})=\frac{1}{2\inum\epsilon}\frac{1}{\lambda_{k}-\bar{\lambda}_{k}+\mathcal{O}(\epsilon^{2})}\prod_{i\ne k}-\frac{\lambda_{k}-\lambda_{i}+\mathcal{O}(\epsilon^{2})}{\lambda_{k}-\bar{\lambda}_{i}+\mathcal{O}(\epsilon^{2})},
\]
and thus
\begin{align*}
\frac{da}{dz}(z_{k})= & \frac{da_{\text{ideal}}}{dz}(z_{k})+\mathcal{O}(\epsilon^{2})\\
= & \frac{du}{dz}(z_{k})\underbrace{\mathcal{B}(z_{k})}_{=0}+u(z_{k})\frac{d\mathcal{B}}{dz}(z_{k})+\mathcal{O}(\epsilon^{2})\\
= & \frac{u(z_{k})}{2\inum\epsilon}\frac{1}{\lambda_{k}-\bar{\lambda}_{k}+\mathcal{O}(\epsilon^{2})}\prod_{i\ne k}-\frac{\lambda_{k}-\lambda_{i}+\mathcal{O}(\epsilon^{2})}{\lambda_{k}-\bar{\lambda}_{i}+\mathcal{O}(\epsilon^{2})}\\
 & +\mathcal{O}(\epsilon^{2}).
\end{align*}
Equation (\ref{eq:asymptotic-norming-constants}) now follows when
the previous formula is plugged into (\ref{eq:discrete-norming-constants}):
\begin{align*}
\tilde{Q}_{k}= & -\frac{b(z_{k})}{2\inum\epsilon z_{k}}\Big/\Big(\frac{u(z_{k})}{2\inum\epsilon}\frac{1}{\lambda_{k}-\bar{\lambda}_{k}+\mathcal{O}(\epsilon^{2})}\\
 & \times\prod_{i\ne k}-\frac{\lambda_{k}-\lambda_{i}+\mathcal{O}(\epsilon^{2})}{\lambda_{k}-\bar{\lambda}_{i}+\mathcal{O}(\epsilon^{2})}+\mathcal{O}(\epsilon^{2})\Big)\\
= & -\frac{b(z_{k})}{z_{k}}\Big/\Big(\frac{u(z_{k})}{1}\frac{1}{\lambda_{k}-\bar{\lambda}_{k}+\mathcal{O}(\epsilon^{2})}\\
 & \times\prod_{i\ne k}-\frac{\lambda_{k}-\lambda_{i}+\mathcal{O}(\epsilon^{2})}{\lambda_{k}-\bar{\lambda}_{i}+\mathcal{O}(\epsilon^{2})}+\mathcal{O}(\epsilon^{2})\Big)\\
\stackrel{\epsilon\to0}{\longrightarrow} & -\frac{B(\lambda_{k})}{U(\lambda_{k})}(\lambda_{k}-\bar{\lambda}_{k})\prod_{i\ne k}-\frac{\lambda_{k}-\bar{\lambda}_{i}}{\lambda_{k}-\lambda_{i}},
\end{align*}
where $U(\lambda)$ and $B(\lambda)$ have been defined in Section
\ref{sub:Many-Samples-Analysis}. 

\begin{figure}
\centering{}\includegraphics[width=0.5\textwidth]{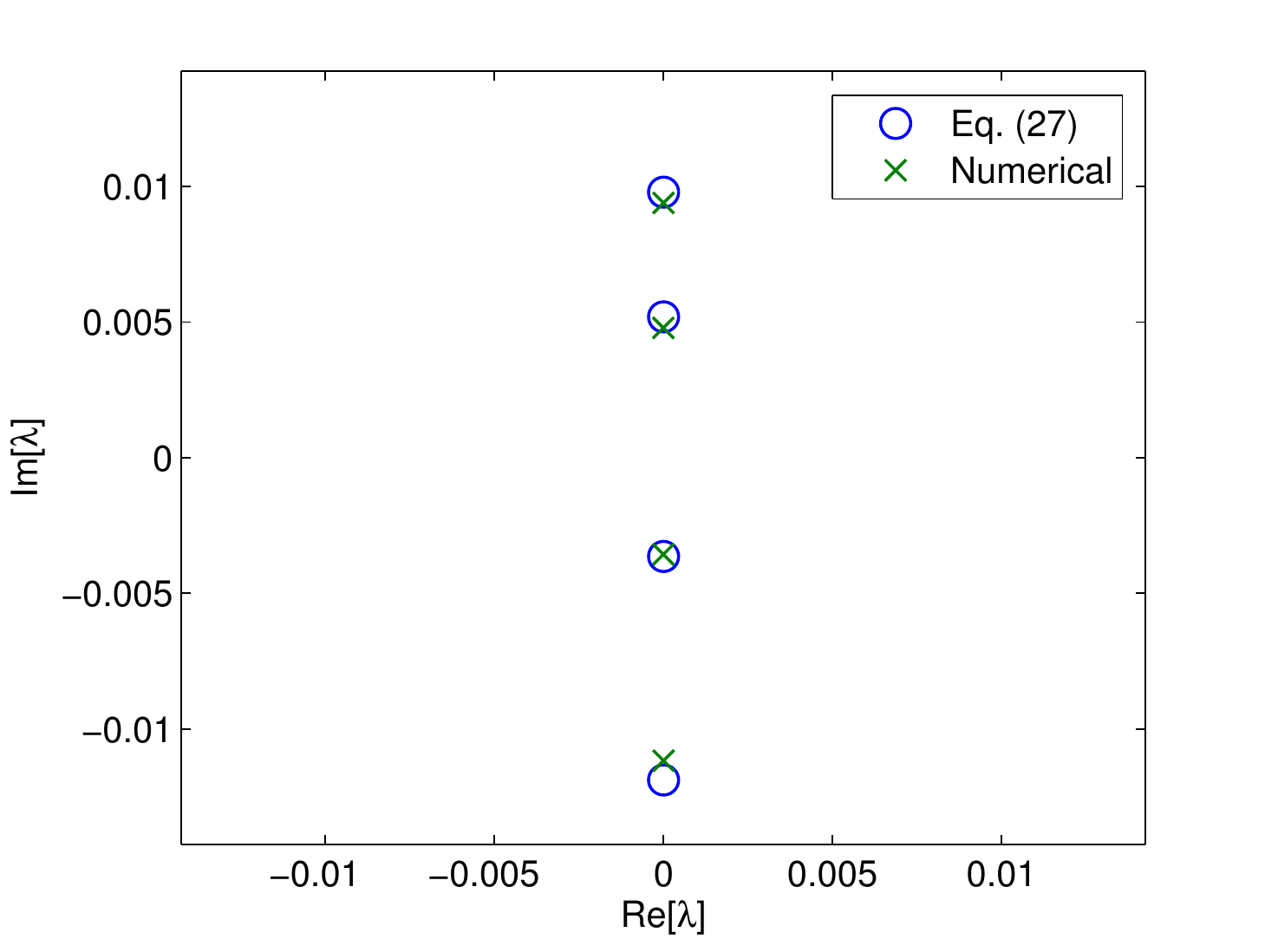}\caption{\label{fig:Norming-constants}Norming constants for the second numerical
experiment}
\end{figure}

We have not yet analyzed the limits $U(\lambda)$ and $B(\lambda)$,
as was already mentioned in Section \ref{sub:Many-Samples-Analysis}.
However, we compared the prediction made by the semi-asymptotic formula
\begin{equation}
\tilde{Q}_{k}\approx-\frac{b(z_{k})}{u(z_{k})}(\lambda_{k}-\bar{\lambda}_{k})\prod_{i\ne k}-\frac{\lambda_{k}-\bar{\lambda}_{i}}{\lambda_{k}-\lambda_{i}}.\label{eq:norming-constants-semi-asymptotic}
\end{equation}
with the exact norming constants that were determined numerically
for the second numerical experiment in order to investigate how fast
the norming constants converge towards their many samples limit. The
results are shown in Figure \ref{fig:Norming-constants}. The values
obtained from (\ref{eq:norming-constants-semi-asymptotic}) match
the true norming constants quite well, but not as well as the roots
match their desired values in Figure \ref{fig:Second-example} (right).
This is consistent with the results reported in \cite[p. 4339]{Yousefi2012b},
were it was found that the norming constants (or \emph{spectral amplitudes})
converge slowly.

\end{document}